\documentclass[conference]{IEEEtran}
\IEEEoverridecommandlockouts
\usepackage{cite}
\usepackage{amsmath,amssymb,amsfonts}
\usepackage{amssymb}
\usepackage{algpseudocode}
\usepackage{amsfonts}
\usepackage{graphicx}
\usepackage{fancyhdr}  %
\usepackage{cases}
\usepackage{textcomp}
\usepackage{extarrows}
\usepackage{algorithm,algpseudocode}

\usepackage{multirow,tabularx}
\usepackage{mathtools}
\usepackage{xcolor}
\usepackage[english]{babel}
\usepackage{amsthm}

\def\BibTeX{{\rm B\kern-.05em{\sc i\kern-.025em b}\kern-.08em
    T\kern-.1667em\lower.7ex\hbox{E}\kern-.125emX}}

\begin{document}

\title{Bidirectional Approximate Message Passing for RIS-Assisted Multi-User MISO Communications}


%
%
%

	\author{
	\IEEEauthorblockN{Li Wei\IEEEauthorrefmark{1}, Chongwen Huang\IEEEauthorrefmark{2}, Qinghua Guo\IEEEauthorrefmark{3}, Zhaoyang Zhang\IEEEauthorrefmark{2}, M\'{e}rouane~Debbah\IEEEauthorrefmark{4} and  Chau~Yuen\IEEEauthorrefmark{1} }
	\IEEEauthorblockA{\IEEEauthorrefmark{1}Singapore University of Technology and Design, 487372 Singapore}
	\IEEEauthorblockA{\IEEEauthorrefmark{2}Department of Information and Electronic Engineering, Zhejiang University }
	\IEEEauthorblockA{\IEEEauthorrefmark{3}School of Electrical, Computer and Telecommunications Engineering, University of Wollongong, Wollongong, Australia}
	\IEEEauthorblockA{\IEEEauthorrefmark{4} Lagrange Mathematics and Computing Research Center, Paris, 75007 France}
	
}

\maketitle

\begin{abstract}
Reconfigurable intelligent surfaces (RISs) have been recently considered as a promising candidate for energy-efficient solutions in future wireless networks. Their dynamic and low-power configuration enables coverage extension, massive connectivity, and low-latency communications. Due to a large number of unknown variables referring to the RIS unit elements and the transmitted signals, channel estimation and signal recovery in RIS-based systems are the ones of the most critical technical challenges. To address this problem, we focus on the RIS-assisted multi-user wireless communication system and present a joint channel estimation and signal recovery algorithm in this paper. Specifically, we propose a bidirectional approximate message passing algorithm that applies the Taylor series expansion and Gaussian approximation to simplify the sum-product algorithm in the formulated problem. Our simulation results show that the proposed algorithm shows the superiority over a state-of-art benchmark method. We also provide insights on the impact of different RIS parameter settings on the proposed algorithms.
\end{abstract}


\section{Introduction}\label{sec:intro}
Reconfigurable Intelligent Surfaces (RISs) is a potential candidate technology for beyond fifth-generation (5G) wireless communications \cite{Akyildiz2018mag,hu2018beyond,huang2019reconfigurable,Marco2019,qingqing2019towards}. A large number of hardware-efficient passive reflecting elements are employed in a RIS to facilitate low-power, energy-efficient, high-speed, massive-connectivity, and low-latency communications  \cite{Akyildiz2018mag,9206044,9136592}. Each unit element can alter the phase of the incoming signal without requiring a dedicated power amplifier that is needed in conventional amplify-and-forward relaying systems \cite{Akyildiz2018mag,9136592,huang2019reconfigurable}. As a result, RISs had gained much attention in recent years. 

The energy efficiency potential of RIS in the scenario of outdoor multi-user multiple input single output communications was analyzed in \cite{huang2019reconfigurable}, while \cite{husha_LIS2} focused on an indoor scenario to illustrate the potential of RIS-based indoor positioning. Recently, a novel passive beamforming and information transfer technique was proposed in \cite{8941126} to enhance primary communications. RIS-assisted communications in the millimeter-wave  and terahertz bands were also lately investigated to deal with limited transmission distance problems  \cite{Akyildiz2018mag}. Orthogonal and non-orthogonal multiple access in RIS-assisted communications were studied in \cite{9133094} as cost-effective solutions for boosting spectrum/energy efficiency. The existing research works have proved the great potential of RISs, however, most of the existing research works focusing on RIS configuration optimization or channel estimation part only, and it is a challenging issue to solve the joint channel estimation and signal recovery problem due to a large number of unknowns, such as a large number of passive elements in RISs and the transmitted signal. 

The existing literature mainly adopts a two-stage approach to estimate channels and transmitted signal separately \cite{9417121}, however, such a method cannot fully explore the characteristics of channels and transmitted signals, and the training overhead is high. For joint estimation and signal recovery, some bilinear message passing algorithms were proposed  \cite{6898015, 8580585,8357527}. In \cite{9103622}, the authors designed a Bayesian method for the effective channel estimation and signal recovery in grant-free non-orthogonal multiple access. In \cite{9006927}, the authors applied a new expectation maximization message passing algorithm combination for joint channel estimation and symbol detection. In \cite{zou2020multi}, the authors proposed a multi-layer algorithm for joint channel estimation and signal detection in two-hop amplify-and-forward relay communication systems with one known channel. However, these approximate message passing (AMP) related algorithms are vulnerable to ill-conditioned measurement matrices that may cause divergence. Thus, some variants to improve the convergence were proposed, such as damping method \cite{6998861} and  AMP with unitary transformation (UTAMP) \cite{9293406,guo2015approximate}. This research provides some insights into the joint channel estimation and signal recovery in RIS-assisted communication systems, which motivates us to explore a new technique to reliably estimate channels and recover signal simultaneously with tolerable training overhead. 

In this paper, we propose a novel joint channel estimation and signal detection algorithm in a RIS-assisted wireless communication system, where a multi-antenna base station (BS) serves multiple single-antenna users. Specifically, we formulate the joint channel estimation and signal recovery as an inference problem that estimates two cascaded channels and the transmitted signal simultaneously. The factor graph and the related sum-product message passing rules of the formulated problem are developed, then we apply the Taylor series expansion and Gaussian approximation to deal with the tricky inference problem.  The proposed bidirectional approximate message passing (BAMP) algorithm is efficient and provides good channel estimation and signal recovery performance. Our extensive simulation results validate the effectiveness of the proposed technique and its favorable performance.  

The remainder of this paper is organized as follows. In Section \ref{sec:format}, the system model is introduced and the estimation problem is formulated. The factor graph and the proposed BAMP algorithm are presented in Section~\ref{sec:channel_est}. Section \ref{sec:simulation} presents the numerical results of the proposed algorithm. Finally, some conclusions are drawn in Section~\ref{sec:conclusion}.

\textit{Notation}: Fonts $a$, $\mathbf{a}$, and $\mathbf{A}$ represent scalars, vectors, and matrices, respectively. We use $\mathbf{A}^T$, $\mathbf{A}^H$, $\mathbf{A}^{-1}$ and $\mathbf{A^\dag}$  to denote the transpose, Hermitian (conjugate transpose), inverse and pseudo-inverse of $ \mathbf{A} $, respectively. The $(m,n)$-th entry of $\mathbf{A}$ is denoted by $a_{mn}$.   $|\cdot|$ and $(\cdot)^*$ denote the modulus and conjugation, respectively. Finally, notation $diag(\mathbf{a})$ represents a diagonal matrix with the entries of $\mathbf{a}$ on its main diagonal.

\section{System Model}\label{sec:format}
In this section, we describe the system model for the considered RIS-empowered wireless communication system.

\begin{figure}\vspace{-2mm}
	\begin{center}
		\centerline{\includegraphics[width=0.5\textwidth]{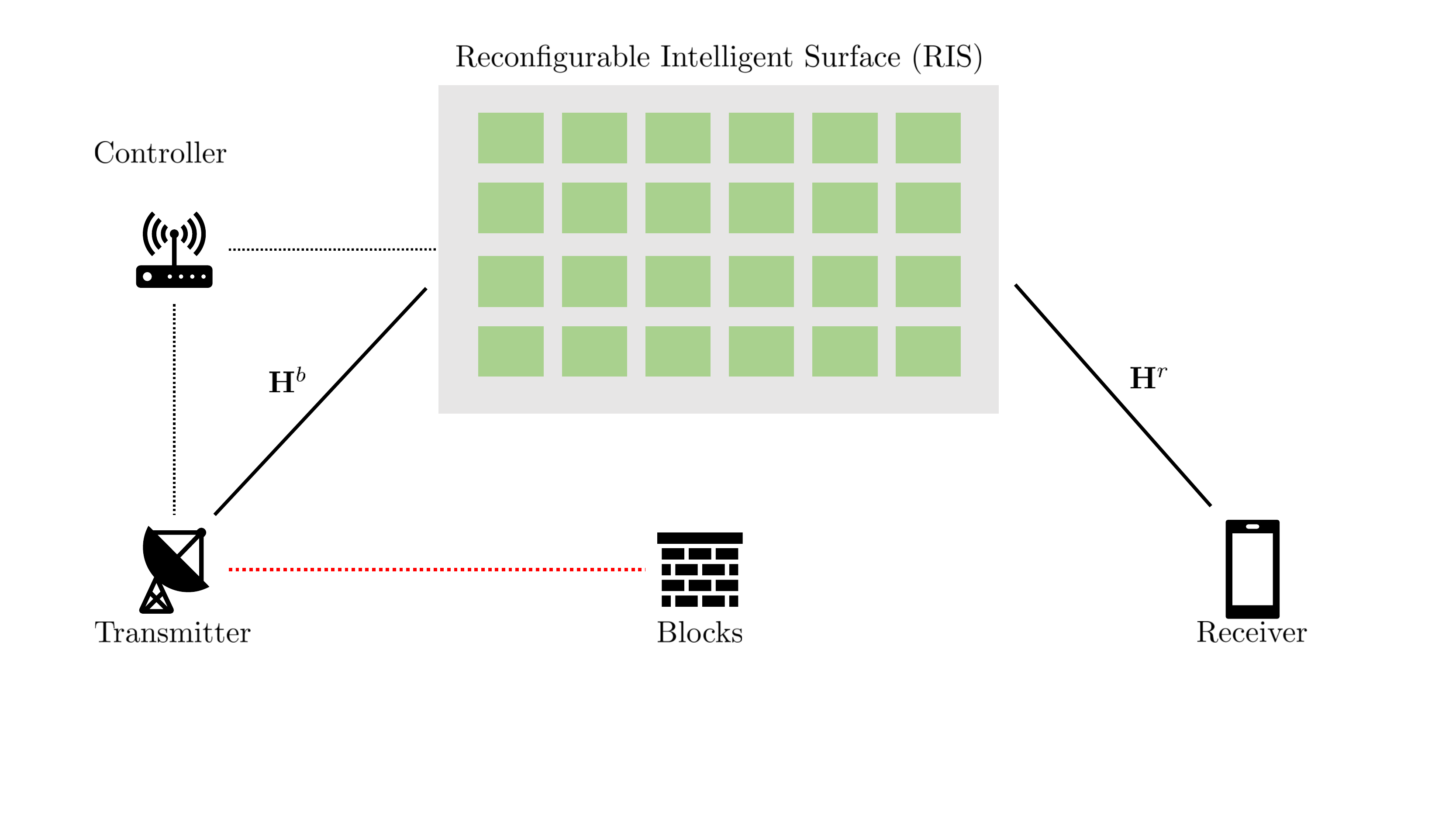}}  \vspace{-0mm}
		\caption{Considered RIS-based wireless communication system consisting of a $M$-antenna BS simultaneously serving in the downlink $K$ single-antenna mobile users.}
		\label{fig:system_model} \vspace{-6mm}
	\end{center}
\end{figure}

We consider the communication between a BS equipped with $M$ antenna elements and $K$ single-antenna mobile users. We assume that this communication is realized via a discrete-element RIS deployed on the facade of a building in the vicinity of the BS side, as illustrated in Fig$.$~\ref{fig:system_model}. The RIS is comprised of $N$ unit cells of equal small size, and each made from metamaterials that are capable of adjusting their reflection coefficients. We assume there is no direct signal path between the BS and  users due to unfavorable propagation conditions, e.g., the presence of large obstacles. The received  signals at all $K$ mobile users for $T$ consecutive time slots can be compactly expressed with $\tilde{\mathbf{Y}} \in\mathbb{C}^{K \times T}$ given by
\begin{equation}\label{equ:YXZ}
	\tilde{\mathbf{Y}} \triangleq\tilde{\mathbf{H}}^{r}  \mathbf{\Phi}   \tilde{\mathbf{H}}^{b} \tilde{\mathbf{X}}+\tilde{\mathbf{W}},
\end{equation}
where diagonal matrix $\mathbf{\Phi}$ is the phase configuration for $N$ RIS unit elements, which is usually chosen from low resolution discrete sets \cite{9110869}; $\tilde{\mathbf{H}}^{b}\in\mathbb{C}^{N\times M}$ and $\tilde{\mathbf{H}}^{r}\in\mathbb{C}^{K\times N}$ denote the channel matrices between RIS and BS, and between all users and RIS, respectively; the matrix  $\tilde{\mathbf{X}}\in\mathbb{C}^{M \times T}$ includes the BS transmitted signal within $T$ time slots; and $\tilde{\mathbf {W}} \in \mathbb{C}^{K \times T}$ is the Additive White Gaussian Noise (AWGN) matrix having zero mean and  variance $N_0$. 
	
In typical cellular configuration, the involved channels are correlated random vectors that are dependent of scattering geometry, however, for uniform linear array with large antenna number at BS, the channels can be represented by sparse matrices in beam domain \cite{6940305,837052,7727995}. Using methods in \cite{6940305,9367220}, the channels in beam domain can be representd as 
\begin{equation}
	\mathbf{H}^b=\tilde{\mathbf{H}}^b \mathbf{F}_1, \mathbf{H}^r=\mathbf{F}_2 \tilde{\mathbf{H}}^r,
\end{equation}
where $\mathbf{F}_1$ denote the $M \times M$ discrete Fourier transform (DFT) matrix, and  $\mathbf{F}_2$ denote the $K \times K$ DFT matrix. Thus, the input-output relationship \eqref{equ:YXZ} can be rewritten as 
\begin{equation}
	\tilde{\mathbf{Y}} \triangleq  \mathbf{F}_2^H {\mathbf{H}}^{r}  \mathbf{\Phi}   {\mathbf{H}}^{b} \mathbf{F}_1^H \tilde{\mathbf{X}}+\tilde{\mathbf{W}} \Rightarrow \mathbf{Y}\triangleq\mathbf{H}^{r}  \mathbf{\Phi}  \mathbf{H}^{b} \mathbf{X} + \mathbf{W}, 
\end{equation}
where $\mathbf{X}=\mathbf{F}_1^H \tilde{\mathbf{X}}$, $\mathbf{Y}=\mathbf{F}_2  \tilde{\mathbf{Y}}$ and $ \mathbf{W}=\mathbf{F}_2  \tilde{\mathbf{W}}$. The beam domain representation yields an equivalent sparse channel estimation in beamspace, which facilitates the joint channel estimation and signal recovery. 

\section{Problem Formulation and Message Passing Algorithm}\label{sec:channel_est}
\subsection{Problem Formulation and Factor Graph Representation}
The focus of this paper is to design an efficient receiver to estimate transmitted signal $\mathbf{X}$ and all involved channels $\mathbf{H}^b$ and $\mathbf{H}^r$. To this end, we formulate a two-layer estimation problem. Specifically, in the first layer, the input is transmitted signal $\mathbf{X}$, and the output is $\mathbf{U}=\mathbf{H}^b \mathbf{X}$. In the second layer, the input is $\mathbf{U}$, and the output is $\mathbf{A}=\mathbf{Q}  \mathbf{U}$, with $\mathbf{Q}=\mathbf{H}^r \mathbf{\Phi}$ that incorporates the unknown channel $\mathbf{H}^r$. The output $\mathbf{A}$ is corrupted by the noise $\mathbf{W}$, which is interpreted as $\mathbf{Y}=\mathbf{A}+\mathbf{W}$. Thus, the joint probability $p(\mathbf{Q},  \mathbf{H}^b,  \mathbf{X},  \mathbf{Y}) $ can be factorized into 
\begin{equation} \label{equ:factor}
		\!p (\!\mathbf{Q}\!, \! \mathbf{H}^b\!, \! \mathbf{X}\!, \!\mathbf{Y}\!) \!\propto\! p(\mathbf{X}) p(\! \mathbf{H}^b \!) \!p(\!\mathbf{U} \!\mid\! \mathbf{H}^b \!\mathbf{X}\!)\! p (\! \mathbf{Q}\!)\!  p  (\!\mathbf{A} \!\mid\!  \mathbf{Q}\! \mathbf{U}\!)\! p \!(\!\mathbf{Y}\! \mid \!\mathbf{A}\!)\!.
\end{equation}
 
\begin{figure*} 
	\begin{center}
		\centerline{\includegraphics[width=0.85\textwidth]{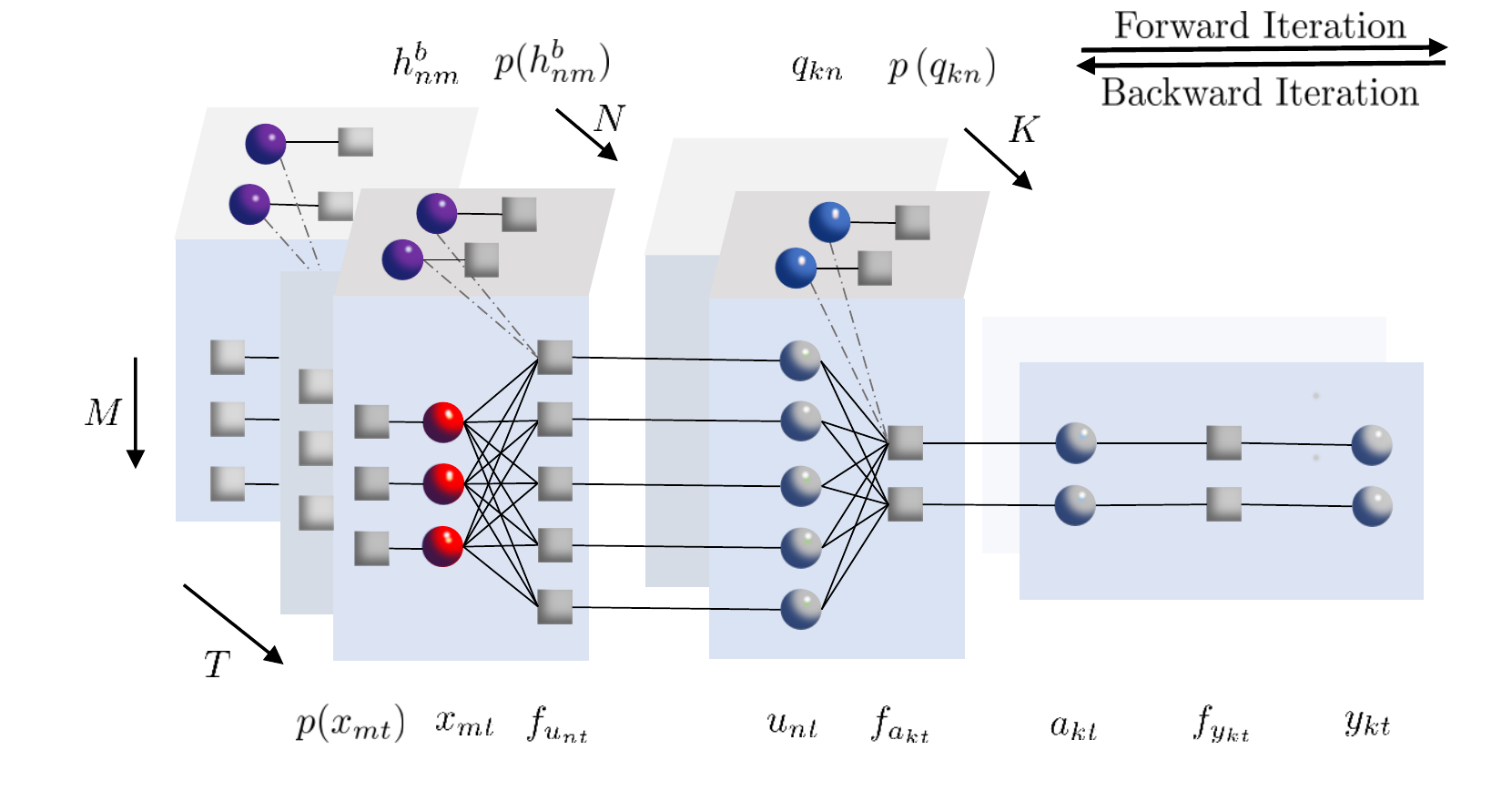} } \vspace{-2mm}
		\caption{The factor graph of the joint channel estimation and signal recovery in RIS-assisted wireless communication systems.}
		\label{fig:factor_graph}
		\vspace{-6mm}
	\end{center}
\end{figure*}  \vspace{-2mm}

The probabilistic structure characterized by \eqref{equ:factor} is illustrated by Fig.~\ref{fig:factor_graph}. The circles represent variables and the squares represent factors. The purple circles denote the variable $h_{nm}^b$; the red circles are variables $x_{mt}$; and the blue circles represent $q_{kn}$. $p(x_{mt}), p(h_{nm}^b)$ and $p(q_{kn})$ are Gaussian priors of the variable $x_{mt}, h_{nm}^b$ and $q_{kn}$, respectively. $f_{u_{nt}}$ is the $(n,t)$-th entry of $p(\mathbf{U} \mid \mathbf{H}^b \mathbf{X})$; $f_{a_{kt}}$ is the $(k,t)$-th entry of $ p(\mathbf{A} \mid  \mathbf{Q} \mathbf{U})$; and $f_{y_{kt}}$ is the $(k,t)$-th element of $p(\mathbf{Y} \mid \mathbf{A})$. As shown in Fig.~\ref{fig:factor_graph}, the message updates bidirectionally, where the message flows from the left to right is termed as forward iteration, and the message flows conversely is backward iteration. The definitions of involved messages are summarized in Table.~\ref{table:messages}

\begin{table}
	\caption{Message definitions in the factor graph.}
	\centering
	\scalebox{0.7}{
		\begin{tabular}{|c|c|c|c|}	\hline 
			\multicolumn{4}{|c|}{\textbf{The first layer}} \\ \hline
			$\mu_{x_{mt} \leftarrow f_{u_{nt}}}  $& message from $f_{u_{nt}}$ to $x_{mt}$  &$	\mu_{x_{mt} \rightarrow f_{u_{nt}}}$ &message from $x_{mt}$ to $f_{u_{nt}}$\\
			\hline   
			$\mu_{f_{u_{nt}} \leftarrow h_{nm}^b}$&message from $h_{nm}^b$ to $f_{u_{nt}}$ &$\mu_{f_{u_{nt}} \rightarrow h_{nm}^b}$& message from $f_{u_{nt}}$ to $h_{nm}^b$  \\
			\hline
			$\mu_{f_{u_{nt}} \leftarrow u_{nt} } $ & message from $u_{nt}$ to $f_{u_{nt}}$& $\mu_{f_{u_{nt}} \rightarrow u_{nt} }$ & message from $f_{u_{nt}}$ to $u_{nt}$ \\
			\hline
			\multicolumn{4}{|c|}{\textbf{The second layer}} \\ \hline
			$\mu_{u_{nt} \leftarrow f_{a_{kt}} }$ &  message from $f_{a_{kt}}$ to $u_{nt}$ &$\mu _{u_{nt} \rightarrow f_{a_{kt}}} $ & message from $u_{nt}$ to $f_{a_{kt}}$ \\
			\hline
			$\mu _{f_{a_{kt}} \rightarrow q_{kn}}$ &  message from $f_{a_{kt}}$ to $q_{kn}$ &  $\mu_{f_{a_{kt}} \leftarrow q_{kn}} $ & message from $q_{kn}$ to $f_{a_{kt}}$ \\
			\hline
			$\mu _{f_{a_{kt}} \rightarrow a_{kt} }$ &  message from $f_{a_{kt}}$ to $a_{kt}$ & $\mu _{f_{a_{kt}} \leftarrow a_{kt} }$ &  message from $a_{kt}$ to $f_{a_{kt}}$ \\
			\hline
			$	\mu_{a_{kt}  \rightarrow  f_{y_{kt}} }$ &  message from $a_{kt}$ to $f_{y_{kt}}$  & $\mu_{f_{y_{kt}}  \rightarrow  y_{kt} }$ &  message from $f_{y_{kt}}$ to $y_{kt}$ \\
			\hline
			$\mu_{a_{kt}  \leftarrow  f_{y_{kt}} }$ &  message from $f_{y_{kt}}$ to $a_{kt}$& & \\
			\hline
	\end{tabular}}
	\label{table:messages}
\end{table}

\subsection{The Proposed BAMP Algorithm}
\label{sec:joint_ce_sr}
Due to the numerous loops and both discrete and continuous-valued variables that are involved in Fig.~\ref{fig:factor_graph}, the exact implementation of the sum-product algorithm is impractical. Thus, we employ the Gaussian approximation and Taylor series expansion to further simplify the messages of loopy belief propagation  for efficient inference, which derives the proposed BAMP algorithm.  

We define the approximate posterior distribution of $\mathbf{U}$ as $\xi_{nt}^{u}$, and 
\begin{equation}
	\begin{aligned}
		&G_{nt}   \left(\mathbb{E}[\xi_{nt}^u],  {\rm Var}[\xi_{nt}^u] \right) \\
		&= \log \! \int \!\mathcal{N} \!\left( \!u_{nt} \!\mid \!\mathbb{E}[\xi_{nt}^u],   {\rm Var}[\xi_{nt}^u] \right)  \mu_{f_{u_{nt}} \leftarrow u_{nt}}  (u_{nt} ) \mathrm{d}{u_{nt}},
	\end{aligned}
\end{equation}
with
\begin{equation}
			\begin{aligned}
		 &Z_{nt} = \sum_{m=1}^{M} { \hat{h}_{nt \leftarrow nm }^b  \hat{x}_{mt \rightarrow nt} },  \\
		&V_{nt}  = \sum_{m=1}^{M}   |\hat{x}_{mt \rightarrow nt}|^2 v_{nt \leftarrow nm}^{b} + |\hat{h}_{nt \leftarrow nm}^b|^2 v_{m t \rightarrow nt}^{x}   \\
		&  \qquad +  v_{nt \leftarrow nm}^{b} v_{m t \rightarrow nt}^{x},  
	\end{aligned}
\end{equation}
where the variable with hat is the means of related messages. 

In the large system limits, the belief $\mu_{x_{mt}} (x_{mt})$ is slightly different from $\mu_{x_{mt  \rightarrow nt}} (x_{mt})$, thus, we further use the means $\hat x_{mt}$ to replace $\hat{x}_{mt \rightarrow nt}$. Besides, the items $ v_{nt \leftarrow nm}^{b} v_{m t \rightarrow nt}^{x} \sim \mathcal{O} (\frac{1}{m})$ and $|\hat{h}_{nt \leftarrow nm}^b|^2 \sim \mathcal{O} (\frac{1}{m})$ are infinitesimal items that can be ignored \cite{zou2020multi}. Thus, we obtain the following using the Taylor series expansion: 
 \begin{equation}
 	\begin{aligned}
 		&\log \mu_{x_{mt} \leftarrow f_{u_{nt}}} (x_{mt})  \\ 
 		& \!\propto\! x_{mt} \left(  \hat{h}_{nm}^b G'_{nt}\left( Z_{nt}, V_{nt} \right) \!-\! |\hat{h}_{nm}^b|^2 \hat{x}_{mt}  G''_{nt}\left( Z_{nt}, V_{nt}\! \right)\!
 		\right) \!
 		\\ &  \! +\! |x_{mt}|^2  \left( \frac{|\hat{h}_{nm}^b|^2}{2} G''_{nt} \left( Z_{nt}, V_{nt} \right) \!+\!   v_{nm}^{b} \dot{G}_{nt} \left( Z_{nt}, V_{nt}\! \right)\! \right)\!,
 	\end{aligned}
 \end{equation}
where $G'_{nt}$ and $G''_{nt}$ are the first and second partial derivatives of $G_{nt}$ w.r.t. the first argument, and $\dot G_{nt}$ is the first derivative w.r.t. its second argument.  We have
\begin{equation}
	\begin{aligned}
		&\tilde{s}_{nt} \triangleq G'_{nt}= \frac{   \tilde{z}_{nt} - Z_{nt}  }   {V_{nt}}, 
	 v^s_{nt} \triangleq -G''_{nt} = - \frac{  \tilde{v}_{nt} - V_{nt}  } {V_{nt}^2},\\
		&\dot G_{nt} = \frac{1}{2} \left[ G_{nt}^{\prime \quad 2} + G''_{nt} \right],
	\end{aligned}
\end{equation}
where $ \frac{   \mathcal{N}  \left( u_{nt} \mid Z_{nt},  V_{nt} \right)    \mu_{f_{u_{nt}} \leftarrow u_{nt}}  (u_{nt} ) }  {\int \mathcal{N}  \left( u_{nt} \!\mid\! Z_{nt},  V_{nt} \right)  \mu_{f_{u_{nt}} \leftarrow u_{nt}}  (u_{nt} ) \mathrm{d}{u_{nt}}}  \!\sim \! \mathcal{N} \left(  u_{nt};  \tilde{z}_{nt} , \tilde{v}_{nt} \right)$.

Thus, the involved messages can be simplified as 
\begin{equation} \label{equ:fu2x}
	\begin{aligned}
		&\!\mu_{x_{mt} \!\leftarrow \!f_{u_{nt}}}\! (x_{mt}) \!\sim\! \mathcal{N} \!\!\left(\!x_{mt} \!\mid\! \frac {\hat{h}_{nm}^b \!\tilde{s}_{nt} \!+\!  \hat{x}_{mt}\! |\hat{h}_{nm}^b|^2 \! v^s_{nt}} {C_x}\! ,\!  \frac{1} {C_x}  \!\right)\!, \\
	   &\!\mu_{\!f_{u_{nt}} \!\rightarrow \!h_{nm}^b} \!(h_{nm}^b)\! \sim\! \mathcal{N} \!\!\left( \!h_{nm}^b \!\mid \! \frac {\hat{x}_{mt} \! \tilde{s}_{nt}\! +\!  \hat{h}_{nm}^b \! |\hat{x}_{mt}^b|^2  \! v^s_{nt}} {C_b}\! ,\!  \frac{1} {C_b} \! \right)\!,
	\end{aligned}
\end{equation}
where $C_x=|\hat{h}_{nm}^b|^2  v^s_{nt} + v^b_{nm} v^s_{nt} - v^b_{nm} |\tilde{s}_{nt}|^2 $, and $C_b=|\hat{x}_{mt}|^2  v^s_{nt} + v^x_{mt} v^s_{nt} - v^x_{mt} |\tilde{s}_{nt}|^2 $.

	The message from $x_{mt}$ to $f_{u_{nt}}$  in $(\ell+1)$-th iteration  is given by
\begin{equation}
	\mu^{\ell+1}_{x_{mt} \rightarrow f_{u_{nt}}} (x_{mt}) \propto p(x_{mt}) \prod_{n' \neq n}^{N} {\mu^{\ell}_{x_{mt} \leftarrow f_{u_{n't}}} (x_{mt})},
\end{equation}
where $p(x_{mt})$ is the $(m,t)$-th element of $p(\mathbf{X})$, and  $p(\mathbf{X})\sim \mathcal{N}(\mathbf{x}^0,\mathbf{v}_x^0)$ is a Gaussian mixture, thus we approximate it to be Gaussian with expectation propagation (EP), which is crucial to achieve the low complexity implementation. Thus, the EP message reads
\begin{equation}
	\begin{aligned}
		\tilde{p}(x_{mt}) \propto \frac{\mu_{x_{mt}}(x_{mt})} {\prod_{n=1}^{N} \mu_{x_{mt}\leftarrow f_{u_{nt}}}(x_{mt})}, 
	\end{aligned}
\end{equation} 
where $\mu_{x_{mt}}(x_{mt})$ is the belief of $x_{mt}$, and the term $\prod_{n=1}^{N} \mu_{x_{mt}\leftarrow f_{u_{nt}}} (x_{mt})$ is the product of Gaussian messages, which can be computed by Gaussian product property.

The beliefs of three unknown variables can be expressed as 
\begin{equation} \label{equ:poste_x}
\begin{aligned}
    &\mu_{x_{mt}}  (x_{mt}) = \tilde{p}(x_{mt}) \prod_{n=1}^{N} \mu_{ x_{mt} \leftarrow f_{u_{nt}}  }  (x_{mt}) \\
    &\propto \tilde{p}(x_{mt}) \mathcal{N} \left( x_{mt} \mid R^{x}_{mt}, \Sigma^{x}_{mt} \right) \sim \mathcal{N} \left( x_{mt} \mid \hat x_{mt}, v^x_{mt}  \right), \\
    &\mu_{h_{nm}^b}  (h_{nm}^b) = p(h_{nm}^b) \prod_{t=1}^{T} \mu_{ f_{u_{nt}}  \rightarrow h_{nm}^b }  (h_{nm}^b) \\
    &\!\propto\! p( h_{nm}^b) \mathcal{N} \left( h_{nm}^b \!\mid \!R^{b}_{nm}, \Sigma^{b}_{nm}  \right) \!\sim \!\mathcal{N} \!\left(\! h_{nm}^b \!\mid \!\hat{h}_{nm}^b, v^b_{nm}  \!\right)\!,
\end{aligned}
\end{equation}
where $ R^{x}_{mt}, \Sigma^{x}_{mt}, R^{b}_{nm}$ and $\Sigma^{b}_{nm}$ can be obtained by Gaussian product property. 

Similarly, the approximate message passing algorihtm can be extended to the second layer, and we have the belif of $q_{kn}$, which  is given by
\begin{equation} \label{equ:poste_r}
	\begin{aligned}
		&\mu_{q_{kn}}  (q_{kn}) = p(q_{kn}) \prod_{t=1}^{T} \mu_{ f_{a_{kt}}  \rightarrow q_{kn} }  (q_{kn}) \\
		&\sim p(q_{kn}) \mathcal{N} \left( q_{kn} \mid R^{q}_{kn}, \Sigma^{q}_{kn}  \right) \sim \mathcal{N} \left( q_{kn} \mid \hat{q}_{kn}, v^q_{kn} \right),
	\end{aligned}
\end{equation}
where $ \mu_{ f_{a_{kt}} \rightarrow q_{kn} }$ can be obtained by a similar way as \eqref{equ:fu2x}, and $R^{q}_{kn}$ and $\Sigma^{q}_{kn}$ can be obtained by Gaussian product property. 
 
With the updated beliefs of all involved variables, the intermediate means and variances in two layers can be further simplified as  
\begin{equation}
	\begin{aligned}
		&Z_{kt}^{\ell} = \sum_{n=1}^{N} { \hat u_{n t \rightarrow kt}^{\ell} \hat q^{\ell}_{kt \leftarrow kn }} \\
		& \!=\! \sum_{n=1}^{N} \! {\left(\hat{u}_{ nt}^{ \ell}- {\hat{q}_{kn}}^{ \ell-1} \tilde{s}_{kt}^{\ell-1} {v_{nt}^{u,\ell}} \right)      
			\left( \hat{q}_{ kn}^{ \ell}- {\hat{u}_{nt}}^{ \ell-1} \tilde{s}_{kt}^{\ell-1} {v_{kn}^{q,\ell}} \right)    } \\
		&\approx \bar Z_{kt}^{\ell} - \tilde{s}_{kt}^{\ell-1} \bar V_{kt}^{\ell},
			\end{aligned}
	\end{equation}
and
		\begin{equation}
			\begin{aligned}
		V_{kt}^{\ell}  & = \!\sum_{n=1}^{N} \!  |\hat u_{n t \rightarrow kt}^{\ell}|^{2} v_{kt \leftarrow kn}^{q,\ell} \!+ \! |\hat q_{ kt \leftarrow  kn }^{\ell}|^2 v_{n t \rightarrow kt}^{u,\ell} \!  \\
			& \qquad+  v_{kt \leftarrow  kn }^{q,\ell} v_{n t \rightarrow kt}^{u,\ell}    \\
		&\approx 	 \bar V_{kt}^{\ell}  +  \sum_{n=1}^{N}  v_{nt}^{u,\ell}  v_{kn }^{q,\ell},
			\end{aligned}
	\end{equation}
where 
\begin{equation}\label{equ:complexity_2nd}
	\begin{aligned}
		&\bar Z_{kt}= \sum_{n=1}^{N}  \hat{u}_{ nt}^{ \ell} \hat{q}_{ kn},\\
		&\bar V_{kt}= \sum_{n=1}^{N} \left(   |{\hat{q}_{kn}} |^2  {v_{nt}^{u}} +  |\hat{u}_{ nt}^{ \ell}|^2  {v_{kn}^{q}}   \right).
	\end{aligned}
\end{equation}

Similarly,
		\begin{equation}
			\begin{aligned}
		Z_{nt}^{\ell} &= \sum_{m=1}^{M} { \hat{h}_{nt \leftarrow nm }^b  \hat{x}_{mt \rightarrow nt} }  \approx \bar Z_{nt}^{\ell} - \tilde{s}_{nt}^{\ell-1} \bar V_{nt}^{\ell}, \\
		V_{nt}^{\ell} &\approx 	 \bar V_{nt}^{\ell}  +  \sum_{m=1}^{M}  v_{mt}^{x,\ell}  v_{nm }^{b,\ell},		
	\end{aligned}
\end{equation}
where
\begin{equation}\label{equ:complexity_1st}
\begin{aligned}
	& \bar Z_{nt}   =  \sum_{m=1}^{M}  \hat{x}_{ mt}  \hat{h}_{ nm}^{b} \\
	 &\bar V_{nt} = \sum_{m=1}^{M}  | \hat{x}_{ mt}|^2 v_{nm}^{b}
	+ |\hat{h}_{ nm}^{ b} |^2 v_{mt}^{x} 
\end{aligned}
\end{equation}

The message between the first layer and the second layer is given by
\begin{equation}
	\mu^{\ell}_{f_{u_{nt}} \leftarrow u_{nt} } (u_{nt})\! =\! \prod_{n=1}^{N} {\mu^{\ell}_{u_{nt} \leftarrow f_{a_{kt} } } } \!\sim \! \mathcal{N}\! \left( \! u_{nt} \!\mid\! R_{nt}, \Sigma_{nt}\! \right),
\end{equation}
which is the product of large number of Gaussian distributions, thus, it can be computed by Gaussian product property.  

The pilots $\mathbf{X}^p \in \mathbb{R}^{M\times T_p}$ are used in the initial iteration, and estimates of $\mathbf{X}$ and $\mathbf{H}^b$ are obtained in the first layer. Then, the output $\hat{\mathbf{U}}=\hat{\mathbf{H}}^b  \hat{\mathbf{X}}$ is considered as input of the inner iteration to update the means of $\mathbf{Q}$.  Although the BAMP two-layer algorithm can estimate all involved unknown channels and signal simultaneously, the divergency issue still exists. This arises from the ill-conditioned matrix $\mathbf{U}$, which is the product of two Gaussian distributed matrices $\mathbf{X}$ and $\mathbf{H}^b$. To further improve the convergence, the damping method is recommended \cite{6998861}.  
 
\subsection{Ambiguities in BAMP} 
The existence of ambiguities undermines the estimation performance of the proposed algorithms, however, there is an inherent ambiguity in recovering the couple $\left( \mathbf{H}^b, \mathbf{X} \right)$ and $\left( \mathbf{Q}, \mathbf{U} \right)$. As a matter of fact, for any invertible unitary matrix $\mathbf{C}_1 \in \mathcal{R}^{M \times M}$ and $\mathbf{C}_2 \in \mathcal{R}^{M \times M}$, the couples $\left( \mathbf{H}^b \mathbf{C}_1, \mathbf{C}_1^{-1} \mathbf{X} \right)$ and $\left( \mathbf{Q}\mathbf{C}_2 , \mathbf{C}_2^{-1} \mathbf{U} \right)$ generate the same values as $\left( \mathbf{H}^b, \mathbf{X} \right)$ and $\left( \mathbf{Q}, \mathbf{U} \right)$.  The ambiguity issue in the proposed algorithm is much more complex than that in the single layer bilinear case, such as BiG-AMP. 

In order to eliminate scaling ambiguities, the first $T_p$ columns of matrix $\mathbf{X}$ and $K_p$ rows of matrix $\mathbf{H}^r$ are assumed to be known \cite{6177982}.  In addition, to remove the phase ambiguity, we adopt the method in \cite{8879620}. In particular, the transmitted signal is designed to be a full-rank matrix. 

\subsection{Computational complexity}
The computational cost of the proposed BAMP algorithm is mainly dominated by componentwise squares of $\mathbf{Q},\mathbf{U},\mathbf{H}^b$ and $\mathbf{X}$. Specifically, in the first layer, the computation complexity is dominated by the computation of $\mathbf{H}^b$ and $\mathbf{X}$ related componentwise squares in \eqref{equ:complexity_1st}, which is $NMT$; and in the second layer, the complexity mainly arises from the computation of $\mathbf{Q}$ and $\mathbf{U}$ related componentwise squares in \eqref{equ:complexity_2nd}, which is $KNT$. Thus, the total computational cost of the proposed algorithm is $\mathcal{O}\left((KNT+NMT)L\right)$ with $L$ being the iteration number.

\section{Simulation Results}\label{sec:simulation}
In this section, we present computer simulation results for the performance of the proposed BAMP algorithm. We have particularly simulated the NMSE using the metrics $\|\mathbf{H}^{b}-\widehat{\mathbf{H}}^{b}\|^{2}\|\mathbf{H}^{b}\|^{-2}$, $\|\mathbf{H}^{r}-\widehat{\mathbf{H}}^{r}\|^{2}\|\mathbf{H}^{r}\|^{-2}$ and $\|\mathbf{X} -\widehat{\mathbf{X}} \|^{2}\|\mathbf{X} \|^{-2}$. The scaling ambiguity of the proposed algorithm has been removed with the aid of the first $K_p$ rows of the channel matrix $\mathbf{H}^{r}$. All normalized mean square error (NMSE) curves were obtained after averaging over $500$ independent Monte Carlo channel realizations. We have used  $L=20$ iterations in all NMSE performance curves. We compare the proposed BAMP with the state-of-art method, the BiGAMP+least squares (LS) method. Specifically, this method consists of two stages: the first stage estimates two channels $\mathbf{H}^r$ and $\mathbf{H}^b$ based on the pilot part $\mathbf{X}_p \in \mathbb{C}^{M \times T_p}$ using BiGAMP; then the data part $\mathbf{X}_d \in \mathbb{C}^{M \times (T-T_p)}$ is estimated based on the obtained channels using the LS in the second stage. 

\begin{figure}  
	\begin{center}
		\includegraphics[width=0.45\textwidth]{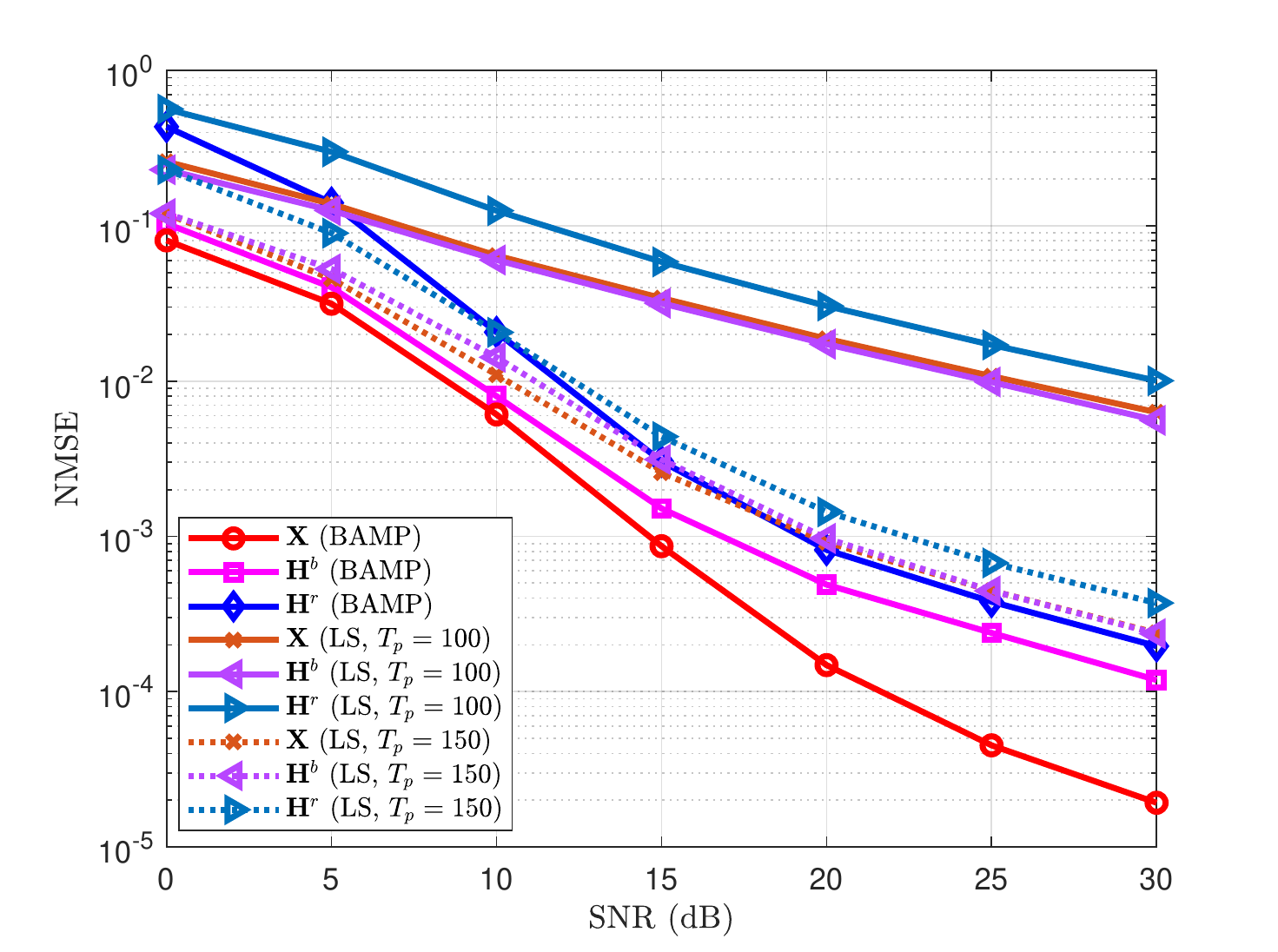}  \vspace{-3mm}
		\caption{NMSE performance comparisons of the BAMP algorithm with baseline method (BiGAMP+LS) versus the SNR in dB for $M=100$, $K=500$, $T=200$ and $N=200$.}
		\label{fig:com_bi_ut} \vspace{-6mm}
	\end{center}
\end{figure}

The NMSE performance comparison of the BAMP two layers algorithm versus the signal-to-noise ratio (SNR) is given in Fig$.$~\ref{fig:com_bi_ut}. The parameter settings are $M=100$, $K=500$, $T=200$, $N=200$ and $K_p=150$, and the damping factor $\beta$ in BAMP method is set $0.3$. In the proposed algorithms, the pilot length is set $T_p=100$.  It can be observed from the figure that the proposed BAMP algorithm  consistently shows great advantages over the benchmark. Specifically, there is  about $18$ dB gap between  the proposed algorithm and the baseline method for the same pilot length $T_p=100$. Even in the unfair setting ($T_p=100$ for the proposed algorithm and  $T_p=150$ for the baseline method), there is about $4$ dB gap between the proposed BAMP and the benchmark in the estimation of $\mathbf{H}^b$, and the gap in the estimation of $\mathbf{X}$ is even larger.  This behavior substantiates the favorable performance of our proposed algorithm. 

\begin{figure} 
	\begin{center}
		\includegraphics[width=0.45\textwidth]{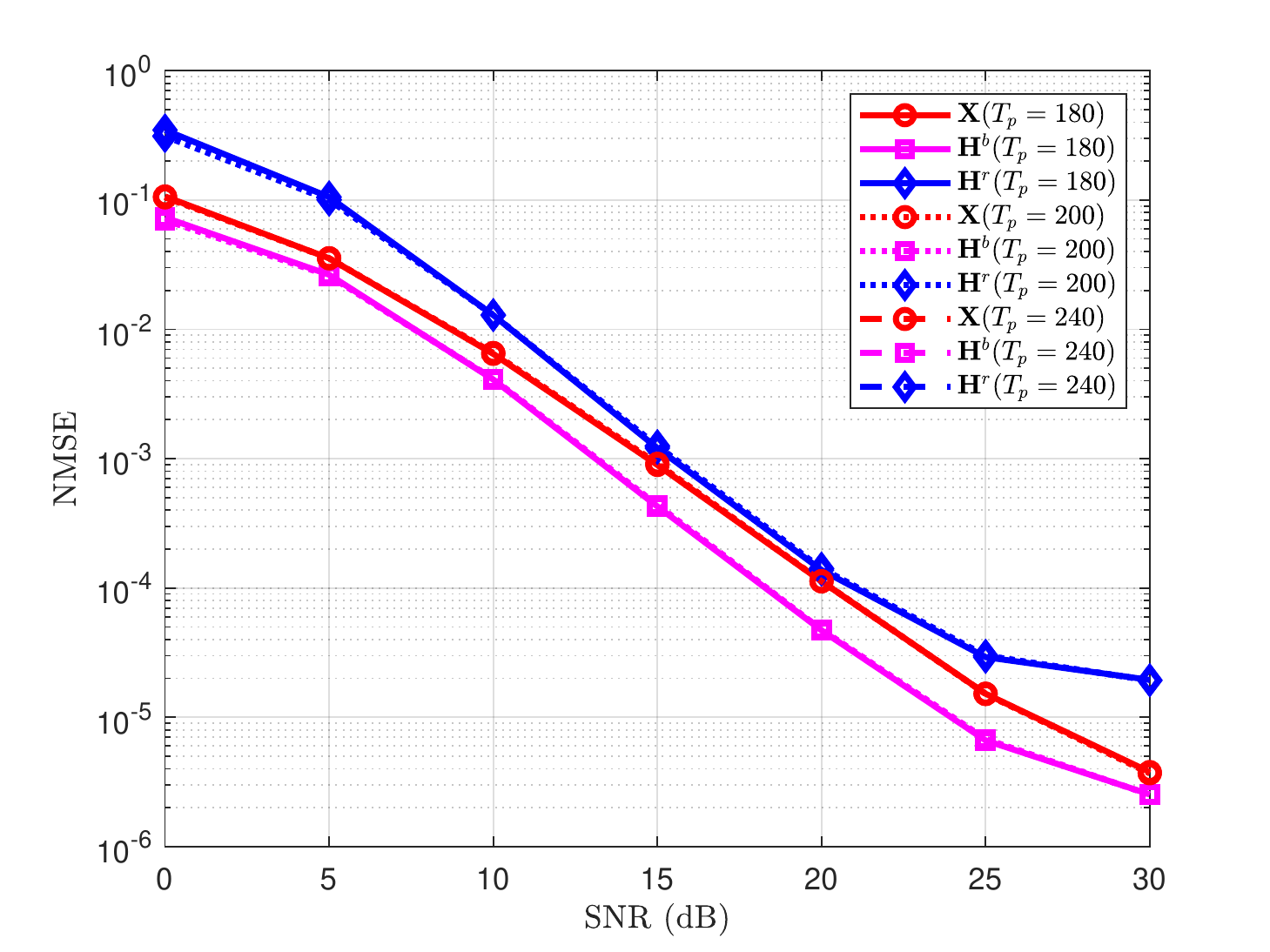}  \vspace{-3mm}
		\caption{NMSE performance comparisons of the BAMP two-layer algorithm versus the SNR in dB for $M=100$, $K=500$, $N=100, T=600$ and various values of $T_p$.}
		\label{fig:bamp_com_tp} \vspace{-6mm}
	\end{center}
\end{figure}

\begin{figure} \vspace{-2mm}
	\begin{center}
		\includegraphics[width=0.45\textwidth]{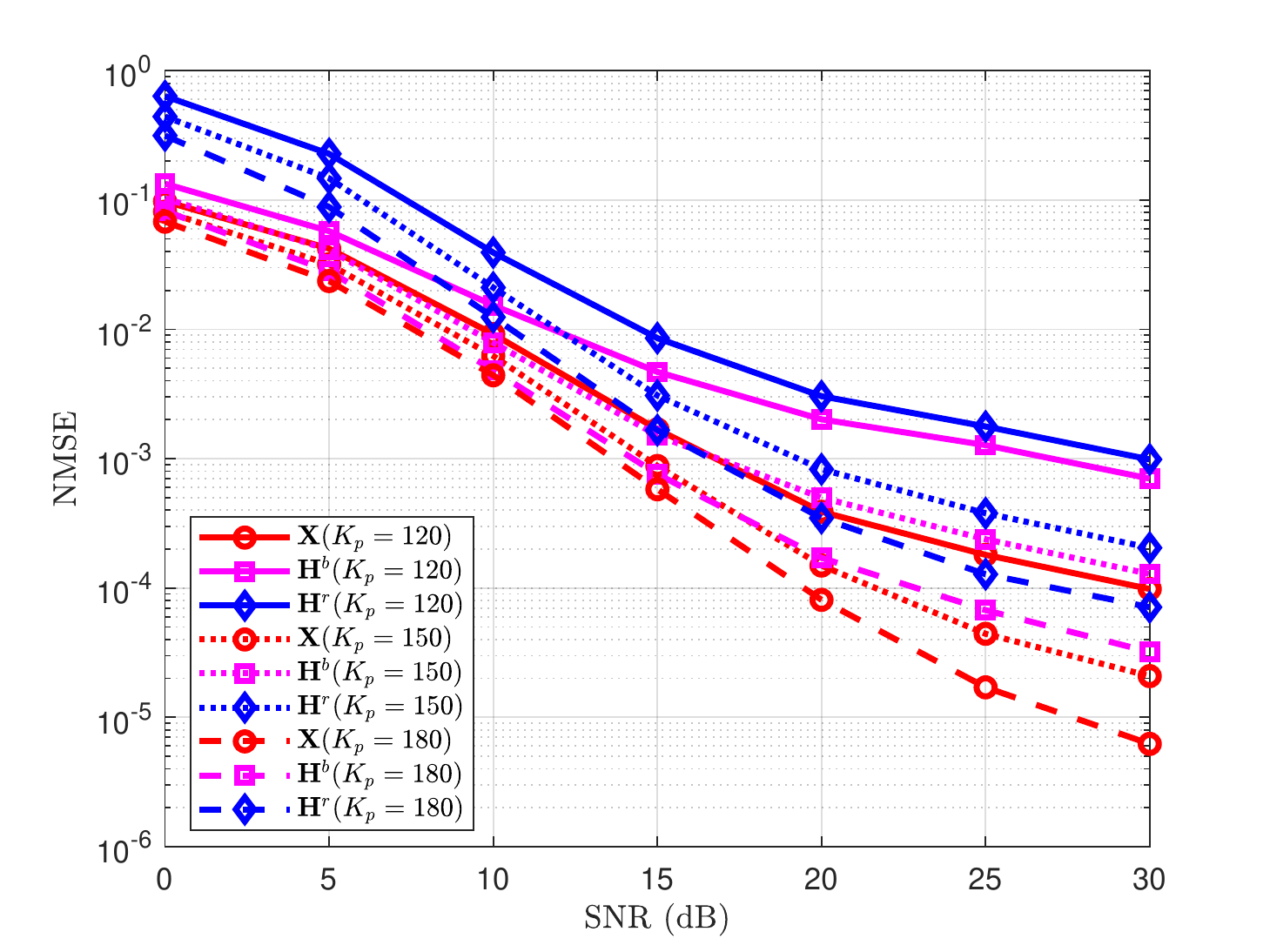}  \vspace{-3mm}
		\caption{NMSE performance comparisons of the BAMP two-layer algorithm versus the SNR in dB for $M=100$, $K=500$, $N=200, T=200$ and various values of $K_p$.}
		\label{fig:bamp_com_kp} \vspace{-6mm}
	\end{center}
\end{figure}

 We evaluate the influence of pilot length in the Fig$.$~\ref{fig:bamp_com_tp} and Fig$.$~\ref{fig:bamp_com_kp}. The minimum number of pilots $T_p$ in $\mathbf{X}$ required for the proposed algrothms is evaluated in Fig$.$~\ref{fig:bamp_com_tp} with  $M=100$, $K=500$, $N=100, T=600$, and $T_p=180, 200, 240$. It can be observed from the figure that the ratio $T_p/T=0.3$ can achieve the similar performance with the case of higher ratio, which means the ratio of pilots in $\mathbf{X}$ with $0.3$ that is enough to achieve the best performance among all cases. In addition, we evaluate the impact of $K_p$ in $\mathbf{H}^r$ in Fig$.$~\ref{fig:bamp_com_kp} with $M=100$, $K=500$, $N=100, T=600$, and $K_p=120, 150, 180$. As shown in figure, the larger $K_p$ increases the whole performance. Taking the estimation of $\mathbf{X}$ as an example, the gap between the case with $K_p=120$ and that with $K_p=150$ is about $9$ dB, and the gap between the case with $K_p=150$ and that with $K_p=180$ reduces to $1.5$ dB. The trend of estimations of $\mathbf{H}^r$ and $\mathbf{H}^r$ is similar, and the performance improvement is even larger than the estimation of $\mathbf{X}$, which substantiates that larger $K_p$ could bring more benefits to the proposed algorithm.  

\begin{figure}  
	\begin{center}
		\includegraphics[width=0.45\textwidth]{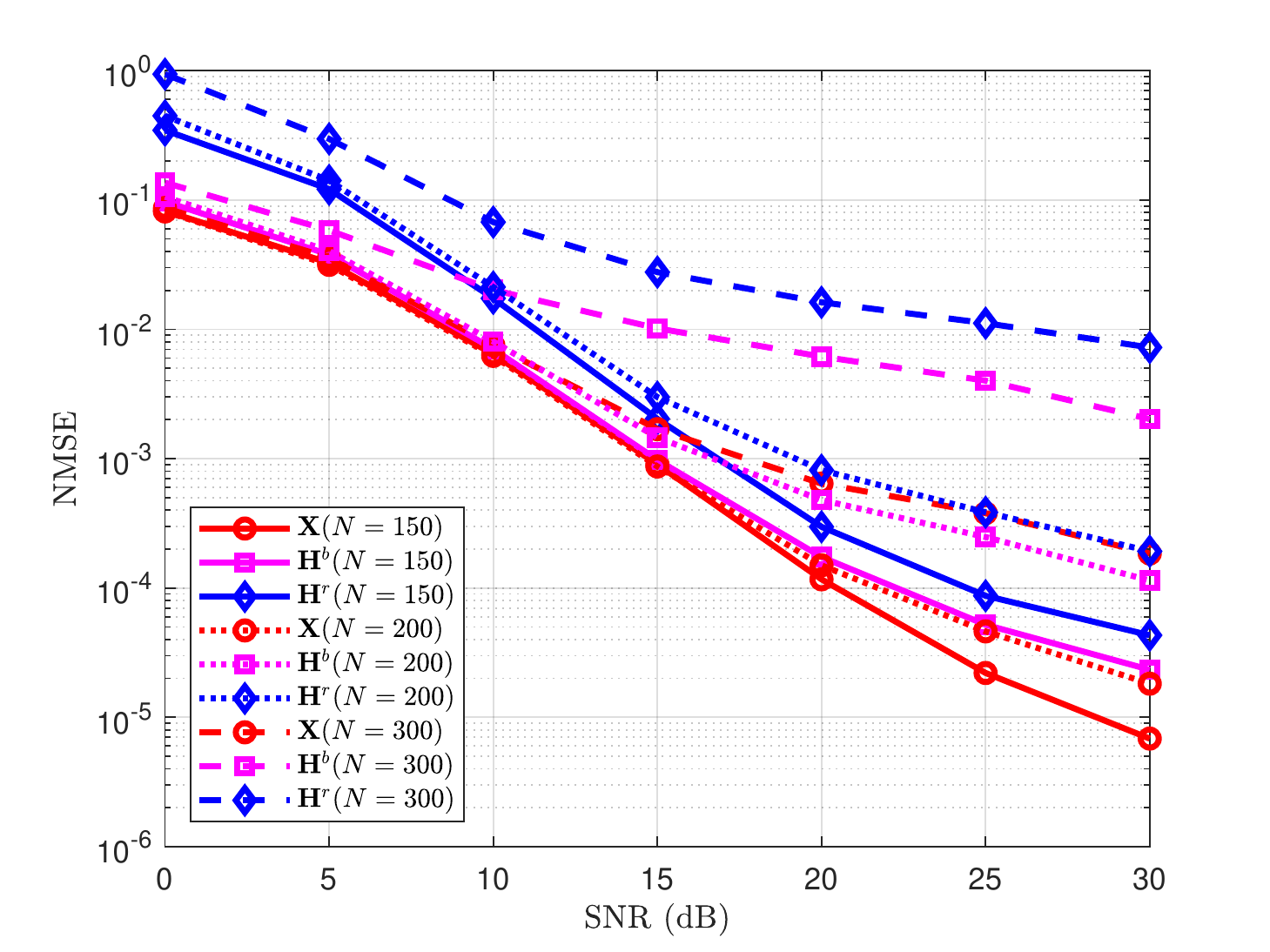}  \vspace{-3mm}
		\caption{NMSE performance comparisons of the BAMP two-layer algorithm versus the SNR in dB for $M=100$, $K=500$, $T=200$ and various values of RIS elements $N$.}
		\label{fig:bamp_com_n} \vspace{-6mm}
	\end{center}
\end{figure}

The performance evaluation of the proposed BAMP two-layer algorithm versus the SNR with various values of RIS elements $N=150, 200$ and $300$ is given in Fig$.$~\ref{fig:bamp_com_n}. The parameter settings are $M=100$, $K=500$, $T=200$, $T_p=100$ and $K_p=150$, and the damping factor $\beta$ is set $0.3$. It is evident that there exists an increasing performance loss when $N$ increases and the gap becomes larger with the increase of SNR, e.g., the gap between the NMSE of $\mathbf{X}$ with $N=150$ and that with $N=200$ is $5$ dB, which is smaller than the gap between cases with $N=150$ and that with $N=300$. In those cases, the number of unknown variables for estimation increases, which results in performance loss.  

\section{Conclusion}\label{sec:conclusion}
In this paper, we proposed a BAMP algorithm for joint channel estimation and signal recovery in RIS-assisted wireless communication systems, which capitalizes on the factor graph and approximate message passing algorithms. All involved channels are estimated and the transmitted signal is recovered through the proposed bidirectional two-layer algorithm. Ambiguities and computational analysis are also presented in this paper. Our simulation results showed that the proposed BAMP algorithm showed superiority over the benchmark scheme even with fewer pilots. In addition, we observed that the pilot length and the number of RIS elements exert a significant effect on our proposed algorithm.


\bibliographystyle{IEEEtran}
\bibliography{strings}
\vspace{12pt}

\end{document}